\documentclass[20pt]{article}
\usepackage{amsmath}
\usepackage{amssymb}
\usepackage{latexsym}
\usepackage{amsmath}
\usepackage{lineno}
\usepackage{graphicx}
\usepackage{float}
\usepackage{subfigure}
\usepackage{geometry}
\usepackage{graphics}
\usepackage{subfigure}
\usepackage{graphicx}
\usepackage{multirow}
\usepackage{booktabs}
\makeatletter

\newcommand{\be}{\begin{equation}}
\newcommand{\ee}{\end{equation}}

\begin{document}
\begin{center}
\large {\bf Corrections to the  thermodynamics of Schwarzschild-Tangherlini black hole and the generalized uncertainty principle}
\end{center}

\begin{center}
Z. W. Feng$^{
 \dagger}$  $\footnote{E-mail:  zwfengphy@163.com}$,
H. L. Li,$^{1,2}$
X. T. Zu,$^{1}$
S. Z. Yang$^{3}$
\end{center}

\begin{center}
\textit{1.School of Physical Electronics, University of Electronic Science and Technology of China, Chengdu, 610054, China\\
2.College of Physics Science and Technology, Shenyang Normal University, Shenyang, 110034, China\\
3.Physics and Space Science College, China West Normal University, Nanchong, 637009, China}
\end{center}

\noindent
{\bf Abstract:} We investigate the thermodynamics of Schwarzschild-Tangherlini black hole in the context of the generalized uncertainty principle.
The corrections to the Hawking temperature, entropy and the heat capacity are obtained via the modified Hamilton-Jacobi equation.
These modifications show that the GUP changes the evolution of Schwarzschild-Tangherlini black hole.
Specially, the GUP effect becomes susceptible when the radius or mass of black hole approach to the order of Planck scale,
it stops radiating and leads to black hole remnant. Meanwhile, the Planck scale remnant can be confirmed through the analysis of the heat capacity.
Those phenomenons imply that the GUP may give a way to solve the information paradox. Besides, we also investigate the possibilities to observe
the black hole at LHC, the results demonstrate that the black hole can not be produced in the recent LHC.

\noindent
{\bf Keywords:} Schwarzschild-Tangherlini black hole; generalized uncertainty principle;  Remnant

\noindent
\section{ Introduction}
One common feature among various quantum gravity theories, such as string theory, loop quantum gravity and non-commutative geometry,
is the existence of a minimum measurable length which can be identified with the order of the Planck scale \cite{ch1,ch2,ch3,ch4}.
This view is also advocated by many Gedanken experiments \cite{ch12}.
The minimum measurable length is especially important since it can be applied into different physical systems and modify many
classical theories \cite{ch7a+,ch13,ch6,ch5,ch14,ch4+,ch5+,ch6+}. One of the most interesting modified theories is called as generalized uncertainty
principle (GUP), which is a generalization of the conventional Heisenberg uncertainty principle (HUP).
It is well known that the uncertainty principle is closely related to the fundamental commutation relation.
Therefore, taking account of the minimum measurable scale, Kempf, Mangano and Mann proposed a modified fundamental commutation relation
\begin{equation}
\label{eq2}
\left[ {x_i ,p_j } \right] = i\hbar \delta _{ij} \left[ {1 + \beta p^2 } \right],
\end{equation}
with the position and momentum operators
\begin{equation}
\label{eq1+}
\begin{array}{*{20}c}
   {x_i  = x_{0i} ,} & {p_j  = p_{0j} \left( {1 + \beta p_0^2 } \right),}  \\
\end{array}
\end{equation}
 where $x_{0i}$  and  $p_{0j} $ satisfying the canonical commutation relations  $\left[ {x_{0i} ,p_{0j} } \right] = i\hbar \delta _{ij}$ \cite{ch11}.
 Through above equations, the most studied form of the GUP is derived as
\begin{equation}
\label{eq1}
\Delta x\Delta p \ge \frac{\hbar }{2}\left[ {1 + \beta \left( {\Delta p} \right)^2 } \right],
\end{equation}
where $\Delta x$ and $\Delta p$ represent uncertainties for position and momentum.
The $\beta  = {{\beta _0 \ell_p^2 } \mathord{\left/ {\vphantom {{\beta _0 \ell_p^2 } {\hbar ^2 }}} \right. \kern-\nulldelimiterspace} {\hbar ^2 }} = {{\beta _0 }
\mathord{\left/ {\vphantom {{\beta _0 } {M_p^2 }}} \right. \kern-\nulldelimiterspace} {M_p^2 }}c^2$,   $\beta_0$ ($\leq 10^{34}$)
is a dimensionless constant, $\ell_p$ and $M_p$ are the Planck length $(\sim10^{-35}m)$ and Planck mass, respectively.
In the HUP framework, the position uncertainty can be measured to an arbitrary small since there is no restriction on the measurement precision of momentum of particles. However, Eq. (\ref{eq1}) implies the GUP existence of minimum measurable length  $\Delta x_{min} \approx \ell_{p} \sqrt{{\beta _{0}}} $.
In the limit $\Delta x \gg \ell_p$, one recovers the HUP $\Delta x\Delta p \ge {\hbar  \mathord{\left/  {\vphantom {\hbar  2}} \right. \kern-\nulldelimiterspace} 2}$.

The implications aspects of GUP have been investigated in many contexts such as modifications of quantum Hall effect \cite{ch15},
neutrino oscillations \cite{ch16}, Landau levels \cite{ch16c+} and cosmology \cite{ch17c+,ch17},  the weak equivalence principle (WEP) \cite{ch18f+}
and Newton¡¯s Law \cite{ch16+,ch16b+}. It should be noted that the GUP has also influence on the thermodynamics of black holes.
In an elegant paper, Adler, Chen and Santiago proposed that the $\Delta p$ and $\Delta x$ of the GUP can be identified as the temperature
and radius of the black hole. With this heuristic method (Hawking temperature-Uncertainty relation), the GUP's impacts on the thermodynamics of Schwarzschild (SC) black hole have been discussed in \cite{ch18}. Their work showed that the modified Hawking temperature is higher than the original case,
and the GUP effect leads to the remnants in the final stages of black hole evaporation. This interesting work has got the attention of people, many other black holes' thermodynamics have been studied with the help of Hawking temperature-Uncertainty relation \cite{ch17+,ch19,ch32b+,ch19+,ch21}.

On the other hand, the thermodynamics of black holes also can be calculated by the tunneling method \cite{ch26a+,ch27a+,ch28a+,ch1x+,ch29a+,ch30a+}.
The tunneling method was first proposed by Parikh and Wilczek for investigating the tunneling behaviors of massless scalar particles \cite{ch26a+}.
Later, this method was extended to study the massive and charged scalar particles tunneling \cite{ch27a+}.
The Hamilton-Jacobi ansatz is another kind of tunneling method \cite{ch28a+,ch1x+,ch29a+}.
With the help of Hamilton-Jacobi ansatz, Kerner and Mann have carefully analyzed the fermion tunneling from black holes \cite{ch30a+}.
So far, the tunneling method plays an important role in studying the black hole radiation, it can effectively help people further understand the properties of black holes, gravity and quantum gravity \cite{ch1e+,ch2e+,ch3e+,ch4e+,ch5e+}.

Combining the GUP with the tunneling method, Nozari and Mehdipour studied the modified tunneling rate of SC black hole \cite{ch1c}. Subsequently,
many more papers on the subject appeared, aiming to investigate the GUP corrected temperature of complicated spacetimes \cite{ch18+,ch27,ch28,ch28+,ch29+,ch20+,ch20b+}.
However, as far as we know, those works are limited to the low dimensional spacetimes. It is well known that the higher dimensions
spacetimes include more physics information, moreover, one of the most exciting signatures is that people may detect the black holes
in the large extra dimensions by using the Large Hadron Collider (LHC) and the Ultrahigh Energy
Cosmic Ray Air Showers (UECRAS) \cite{ch22,ch22+,ch23+,ch23,ch24,ch25,ch3x3}. In other word, the large extra dimensions has opened
up new doors of research in black holes and quantum gravity. Therefore, in this paper, we will investigate the GUP corrected
thermodynamics of Schwarzschild -Tangherlini (ST) black hole by using the quantum tunneling method.
The ST black hole is a typical higher dimensional black hole, people can get many new solutions of higher dimensional spacetimes via the ST metrics.
In \cite{ch26}, the authors showed that the ST black hole is a good approximation to a compactified spacetime when the compact dimension's size
is much larger than the black hole's size. Thus, the ST black hole is a good tool for researching the distorted compactified spacetime.
Based on the above arguments, we think the GUP corrected thermodynamics of ST black hole are worth to be studied.
 By utilizing the tunneling method and the GUP, we find that the modified temperature is lower than the original case.
 Meanwhile, it is also in contrast to the earlier findings, which are analyzed by the Hawking temperature-Uncertainty relation \cite{ch18,ch23}.
 When the mass of the ST black hole reaches the order of Planck scale, the GUP corrected thermodynamics decreases to zero.
 This in turn prevents black hole from evaporating completely and leads to the remnant of ST black hole.

The outline of the paper is as follows. In section 2, incorporating GUP, we derive the modified Hamilton-Jacobi equations in the curved spacetime via WKB
approximation. In Section 3, the tunneling radiation of particles from the ST black hole is addressed.
In section 4, due to the GUP corrected temperature, we analysis the remnants of ST black hole. In section 5, we investigate the minimum black hole energy to form black hole in the LHC. The last section is devoted to our conclusion.

\section{Modified Hamilton-Jacobi Equations}
In this section, we will derive the modified Hamilton-Jacobi equations from the generalized Klein-Gordon equation and generalized Dirac equation. Based on momentum operators of Eq. (\ref{eq1+}), the square of momentum takes the form \cite{ch27,ch28}
\begin{equation}
\label{eq3}
p^2  = p_i p^i  \simeq  - \hbar ^2 \left[ {1 - 2\beta \hbar ^2 \left( {\partial _j \partial ^j } \right)} \right]\left( {\partial _i \partial ^i } \right).
\end{equation}
It is note that the higher order terms $\mathcal{O}(\beta)$  in above equation are ignored.
Adopting the effects of generalized frequency $\bar \omega  = E\left( {1 - \beta E^2 } \right)$  and mass shell condition,
the generalized expression of energy is \cite{ch29}
\begin{equation}
\label{eq4}
\bar E = E\left[ {1 - \beta \left( {p^2  + m^2 } \right)} \right],
\end{equation}
where energy operator is defended as  $E = i\hbar \partial _t$. Therefore, the original Klein-Gordon equation in the curved spacetime is given by
\begin{equation}
\label{eq5}
\left[ {\left( {i\hbar } \right)^2 D^\mu  D_\mu   + m^2 } \right]\Psi  = 0,
\end{equation}
where  $D_\mu =\nabla _\mu   + {{ieA_\mu  } \mathord{\left/ {\vphantom {{ieA_\mu  } \hbar }} \right. \kern-\nulldelimiterspace} \hbar }$ with the geometrically covariant derivative $\nabla _\mu$, the $m$  and $e$  denote the mass and charge of particles, $A^\mu$  is the electromagnetic potential of spacetime, respectively. In order to get the generalized Klein-Gordon equation, Eq. (\ref{eq5}) should be rewritten as
\begin{equation}
\label{eq6}
{ - \left( {i\hbar } \right)^2 \left( {\partial _t  + \frac{i}{\hbar }eA_t } \right)\left( {\partial ^t  + \frac{i}{\hbar }eA^t } \right)\Psi  = \left[ {\left( {i\hbar } \right)^2 \left( {\partial _k  + \frac{i}{\hbar }eA_k } \right)\left( {\partial ^k  + \frac{i}{\hbar }eA^k } \right) + m^2 } \right]\Psi ,}
\end{equation}
where $k = 1,2,3 \cdots$ represent the spatial coordinates. In above equation, the relation $\nabla _\mu   = \partial _\mu $ has been used. The right hand of Eq. (\ref{eq6}) is related to the energy. Inserting the Eq. (\ref{eq3}) and Eq. (\ref{eq4}) into above equation, one can generalizes the original Klein-Gordon equation to following form
\begin{equation}
\label{eq7}
{ - \left( {i\hbar } \right)^2 \left( {\partial _t  + \frac{i}{\hbar }eA_t } \right)\left( {\partial ^t  + \frac{i}{\hbar }eA^t } \right)\Psi  = \left[ {\left( {i\hbar } \right)^2 \left( {\partial _k  + \frac{i}{\hbar }eA_k } \right)\left( {\partial ^k  + \frac{i}{\hbar }eA^k } \right) + m^2 } \right]\left[ {1 - \beta \left( {p^2  + m^2 } \right)} \right]^2 \Psi .}
\end{equation}
 The wave function of generalized Klein-Gordon equation Eq. (\ref{eq7}) can be expressed as  $\Psi  = \exp [{{iS(t,k)} \mathord{\left/ {\vphantom {{iS(t,k)} \hbar }} \right. \kern-\nulldelimiterspace} \hbar }]$, where $S\left( {t,k}\right)$ is the action of the scalar particle. Substituting the wave function into Eq. (\ref{eq7}) and using the WKB approximation, the modified Hamilton-Jacobi equation for scalar particle is gotten as
\begin{equation}
\label{eq8}
{g^{00} \left( {\partial _0 S + eA_0 } \right)^2  + \left[ {g^{kk} \left( {\partial _k S + eA_k } \right)^2  + m^2 } \right]\left\{ {1 - 2\beta \left[ {g^{jj} \left( {\partial _j S} \right)^2  + m^2 } \right]} \right\} = 0.}
\end{equation}

It is well known that the original Dirac equation can be expressed as
 $ - i\gamma ^t \nabla _t \Psi  = (i\gamma ^k \nabla _k  + {m \mathord{\left/  {\vphantom {m \hbar }} \right.  \kern-\nulldelimiterspace} \hbar })\Psi$
 with $\nabla _k   = \partial _k   + \Omega _k   + {{ieA_k  } \mathord{\left/ {\vphantom {{ieA_k  } \hbar }} \right. \kern-\nulldelimiterspace} \hbar }$,
 where the left hand is related to the energy. According to the method in \cite{ch27,ch28},
putting the generalized expression of energy Eq. (\ref{eq3}) and Eq. (\ref{eq4}) into the original Dirac equation,
one yields the generalized Dirac equation in the curved spacetime
\begin{equation}
 - i\gamma ^t \nabla _t \Psi  = \left( {i\gamma ^k \nabla _k  + {m \mathord{\left/
 {\vphantom {m \hbar }} \right.
 \kern-\nulldelimiterspace} \hbar }} \right)\Upsilon \left( \beta  \right)\Psi,
 \label{eq9}
\end{equation}
where $\Upsilon \left( \beta  \right) = 1 - \beta \left( {p^2  + m^2 } \right)$. Since the $t-t$ component of Eq. (\ref{eq9}) is related to the energy, hence it did not be corrected by the GUP term $\Upsilon \left( \beta  \right)$, this leads the Eq. (\ref{eq9}) is different from the generalized Dirac equation $ - i\gamma ^0 \partial _0 \Psi  = (i\gamma ^i \nabla _i  + i\gamma ^t \Omega _t  + {{ieA_t } \mathord{\left/  {\vphantom {{ieA_t } \hbar }} \right.  \kern-\nulldelimiterspace} \hbar } + {m \mathord{\left/  {\vphantom {m \hbar }} \right.  \kern-\nulldelimiterspace} \hbar })\Upsilon \left( \beta  \right)\Psi $ in \cite{ch27,ch28}. Then, multiplying $ - i\gamma ^t \nabla _t  - [i\gamma ^n \nabla _n  - {m \mathord{\left/ {\vphantom {m \hbar }} \right.
 \kern-\nulldelimiterspace} \hbar }]\Upsilon \left( \beta  \right)$ by Eq. (\ref{eq9}), the generalized Dirac equation can be written as
\begin{equation}
\label{eq10}
\begin{array}{l}
 \left\{ { - \left( {\gamma ^t \nabla _t } \right)^2  - \gamma ^t \nabla _t \gamma ^n \nabla _n \Upsilon \left( \beta  \right) - \gamma ^k \nabla _k \gamma ^t \nabla _t \Upsilon \left( \beta  \right)} \right. +  \left. {\left[ {i\left( {\gamma ^n \nabla _n  - \gamma ^k \nabla _k } \right)\frac{m}{\hbar } - \gamma ^k \nabla _k \gamma ^n \nabla _n  - \left( {\frac{m}{\hbar }} \right)^2 } \right]\Upsilon \left( \beta  \right)^2 } \right\} \\
 \\
  \times \Psi  = 0. \\
 \end{array}
\end{equation}
When assuming  $k=n$, above equation becomes to
\begin{equation}
\label{eq11}
{\left\{ { - \frac{{\left\{ {\gamma ^t ,\gamma ^t } \right\}}}{2}\nabla _t^2  - \left[ {\frac{{\left\{ {\gamma ^k ,\gamma ^k } \right\}}}{2}\nabla _k^2  + \left( {\frac{m}{\hbar }} \right)^2 } \right]\Upsilon \left( \beta  \right)^2 } \right\}\Psi  = 0.}
\end{equation}
In order to derive the modified Hamilton-Jacobi equation from Eq. (\ref{eq11}), the wave function of generalized Dirac equation takes on the form
\begin{equation}
\label{eq12}
\Psi  = \xi \left( {t,k} \right)\exp \left[ {{{iS\left( {t,k} \right)} \mathord{\left/
 {\vphantom {{iS\left( {t,k} \right)} \hbar }} \right.
 \kern-\nulldelimiterspace} \hbar }} \right],
\end{equation}
where  $\xi \left( {t,k} \right)$ is a vector function of the spacetime. Denoting  $t=0$, the anti-commute relation gamma matrices obey $\left\{ {\gamma ^0 ,\gamma ^k } \right\} = 0$ ,  $\left\{ {\gamma ^k ,\gamma ^k } \right\} = 2g^{kk} I$ and $\left\{ {\gamma ^0 ,\gamma ^0 } \right\} = 2g^{00} I$. Substituting the gamma matrices anti-commutation relations and Eq. (\ref{eq12}) into Eq. (\ref{eq11}), the resulting equations to leading order in $\beta$ is
\begin{equation}
\label{eq15+}
{\left\{ {g^{00} \left( {\partial _t S + eA_t } \right)^2  + \left[ {g^{kk} \left( {\partial _k S + eA_k } \right)^2  + m^2 } \right]\left. {\left\{ {1 - 2\beta \left[ {g^{jj} \left( {\partial _j S} \right)^2  + m^2 } \right]} \right\}} \right\}\xi \left( {t,k} \right) = 0.} \right.}
\end{equation}
Eq. (\ref{eq15+}) for the coefficient will has a non-trivial solution if and only if the determinant vanishes, that is
\begin{equation}
\label{eq15}
{{\rm{Det}}\left\{ {g^{00} \left( {\partial _t S - eA_t } \right)^2  + \left[ {g^{kk} \left( {\partial _k S + eA_k } \right)^2  + m^2 } \right]\left\{ {1 - 2\beta \left[ {g^{jj} \left( {\partial _j S} \right)^2  + m^2 } \right]} \right\}} \right\} = 0.}
\end{equation}
When keeping the leading order term of $\beta$, the modified Hamilton-Jacobi equation for fermion is directly obtained as
\begin{equation}
\label{eq16}
{g^{00} \left( {\partial _0 S + eA_0 } \right)^2  + \left[ {g^{kk} \left( {\partial _k S + eA_k } \right)^2  + m^2 } \right]\left\{ {1 - 2\beta \left[ {g^{jj} \left( {\partial _j S} \right)^2  + m^2 } \right]} \right\} = 0.}
\end{equation}
Comparing Eq. (\ref{eq8}) with Eq. (\ref{eq16}), it is clear that the modified Hamilton-Jacobi equations for scalar particle and fermions are the similar.
In \cite{ch3e+,ch20+,ch31d+}, the authors derived the Hamilton-Jacobi from the Rarita-Schwinger equation, the Maxwell's equations and the gravitational wave equation, they indicated that the Hamilton-Jacobi equation can describe the behavior of particles with any spin in the curve spacetime.
As we know, the Hamilton-Jacobi ansatz can greatly simplify the workload in the research of black hole radiation.
Especially for fermion tunneling case, people do not need construct the tetrads and gamma matrices with the help of the Hamilton-Jacobi equation.
Adopting the modified Hamilton-Jacobi equation, the tunneling radiation of ST black hole will be studied in the next section.

\section{Quantum Tunneling from ST black hole}
To begin with, we need make a few remarks about the ST black hole. In Ref.~\cite{ch32}, the author added extra compact spatial dimensions into a static spherically symmetric spacetime, and obtained the line element of ST black hole
\begin{equation}
\label{eq17}
ds^2  =  - f\left( r \right)dt^2  + f\left( r \right)^{ - 1} dr^2  + r^2 d\Omega _{D - 2}^2 ,
\end{equation}
where $f\left( r \right) = 1 - \left( {{{r_H } \mathord{\left/ {\vphantom {{r_H } r}} \right. \kern-\nulldelimiterspace} r}} \right)^{D - 3}$,  $d\Omega _{D - 2}^2$ is a metric on a unit $D-2$  dimensional sphere, it covered by the original angular coordinates  $\theta _1 ,\theta _2 ,\theta _{3}, \cdots, \theta _{D - 2}$.  $r_H$ is the event horizon of ST black hole, which characterized by the mass $M$\begin{equation}
\label{eq18}
r_H  = \left[ {\frac{{16GM}}{{\left( {D - 2} \right)\varpi _{D - 2} }}} \right]^{\frac{1}{{D - 3}}}  = \frac{1}{{\sqrt \pi  }}\left[ {\frac{{8M\Gamma \left( {\frac{{D - 1}}{2}} \right)}}{{D - 2M_P^{D - 2} }}} \right]^{\frac{1}{{D - 3}}}
\end{equation}
where $G = {1 \mathord{\left/ {\vphantom {1 {M_P^{D - 2} }}} \right. \kern-\nulldelimiterspace} {M_P^{D - 2} }}$ is the $D-2$ dimensional Newton constant  and the volume of the unit $D-2$ dimensional sphere as  $\varpi _{D - 2}  = {{2\pi ^{\frac{{D - 1}}{2}} } \mathord{\left/ {\vphantom {{2\pi ^{\frac{{D - 1}}{2}} } {\Gamma \left( {\frac{{D - 1}}{2}} \right)}}} \right.  \kern-\nulldelimiterspace} {\Gamma \left( {\frac{{D - 1}}{2}} \right)}}$ \cite{ch22}.

Next, we will calculate the quantum tunneling from ST black hole. Inserting the inverse metric of ST black hole into  the modified Hamilton-Jacobi equation, one has
\begin{equation}
\label{eq20}
\begin{array}{l}
 f^{ - 1} \left( {\partial _t S} \right)^2  - \left[ {f\left( r \right)\left( {\partial _r S} \right)^2  + \left( {g^{\theta _1 \theta _1 } } \right)\left( {\partial _{\theta _1 } S} \right)^2  + \left( {g^{\theta _2 \theta _2 } } \right)\left( {\partial _{\theta _2 } S} \right)^2  +  \cdots  + \left( {g^{\theta _{D - 2} \theta _{D - 2} } } \right)\left( {\partial _{\theta _{D - 2} } S} \right)^2 } \right. \\
 \\
 \left. { + m^2 } \right]\left\{ {1 - 2\beta \left[ {f\left( r \right)\left( {\partial _r S} \right)^2  + \left( {g^{\theta _1 \theta _1 } } \right)\left( {\partial _{\theta _1 } S} \right)^2  + \left( {g^{\theta _2 \theta _2 } } \right)\left( {\partial _{\theta _2 } S} \right)^2  +  \cdots  + \left( {g^{\theta _{D - 2} \theta _{D - 2} } } \right)\left( {\partial _{\theta _{D - 2} } S} \right)^2 } \right.} \right. \\
 \\
 \left. {\left. { + m^2 } \right]} \right\} = 0. \\
 \end{array}
\end{equation}
where $\lambda$ is a constant. First, focusing on the Eq. (\ref{eq22}), in \cite{ch20+}, the author showed that the magnitude of the  particles' angular momentum  can be expressed in the terms of $\partial _{\theta _1 } \Theta$, $\partial _{\theta _2 } \Theta$,$ \cdots \partial _{\theta _{D - 2} } \Theta$, that is
\begin{equation}
\label{eq21}
\begin{array}{*{20}c}
   {f\left( r \right)\left( {\partial _r W} \right)^2 \left[ {2\beta f^{ - 1} \left( r \right)\left( {\partial _r W} \right)^2  - 1} \right] + \left[ {f\left( r \right)\left( {\partial _r W} \right)^2  + \left( {g^{\theta _1 \theta _1 } } \right)\left( {\partial _{\theta _1 } \Theta } \right)^2 } \right. + \left( {g^{\theta _2 \theta _2 } } \right)}  \\
   {}  \\
   {\left. { \times \left( {\partial _{\theta _2 } \Theta } \right)^2  +  \cdots  + \left( {g^{\theta _{D - 2} \theta _{D - 2} } } \right)\left( {\partial _{\theta _{D - 2} } S} \right)^2 } \right] \left[ {4\beta f^{ - 1} \left( r \right)\left( {\partial _r W} \right)^2  - 1} \right] + \omega ^2 f^{ - 1} \left( r \right) =  - \lambda ,}  \\
\end{array}
\end{equation}
\begin{equation}
\label{eq22}
{{\rm{2}}\beta \left[ {\left( {g^{\theta _1 \theta _1 } } \right)\left( {\partial _{\theta _1 } \Theta } \right)^2  + \left( {g^{\theta _2 \theta _2 } } \right)\left( {\partial _{\theta _2 } \Theta } \right)^2 \left. { +  \cdots  + \left( {g^{\theta _{D - 2} \theta _{D - 2} } } \right)\left( {\partial _{\theta _{D - 2} } \Theta } \right)^2 } \right]^{\rm{2}} {\rm{ = }}\lambda .} \right.}
\end{equation}
where $\lambda$ is a constant. First, focusing on the Eq. (\ref{eq22}), in \cite{ch20+}, the author showed that the magnitude of the  particles' angular momentum  can be expressed in the terms of $\partial _{\theta _1 } \Theta$, $\partial _{\theta _2 } \Theta$,$ \cdots \partial _{\theta _{D - 2} } \Theta$, that is
\begin{equation}
\label{eq22k}
 \left( {g^{\theta _1 \theta _1 } } \right)\left( {\partial _{\theta _1 } \Theta } \right)^2  + \left( {g^{\theta _2 \theta _2 } } \right)\left( {\partial _{\theta _2 } \Theta } \right)^2  + \cdots  + \left( {g^{\theta _{D - 2} \theta _{D - 2} } } \right)\left( {\partial _{D - 2} \Theta } \right)^2  = \mathcal{L}^2,  \\
\end{equation}
According to Eq. (\ref{eq22k}), one can written Eq. (\ref{eq22}) as
\begin{equation}
\label{eq23}
2\left( {{\cal L}^2 } \right)^2  = {\lambda  \mathord{\left/
 {\vphantom {\lambda  \beta }} \right.
 \kern-\nulldelimiterspace} \beta }.
\end{equation}
In above equation indicates that the constant $\lambda$ is related to the angular momentum of the emitted particle.  With the help of Eq. (\ref{eq22k}) and Eq. (\ref{eq23}), the Eq. (\ref{eq21}) becomes to
\begin{equation}
\label{eq24}
P_4 \left( {\partial _r W} \right)^4  + P_2 \left( {\partial _r W} \right)^2  + P_0  = 0,
\end{equation}
where   $P_4  = 2\beta f\left( r \right)^2$,   $P_2  = \left( {4m^2 \beta  - 1} \right)f\left( r \right)$ and   $P_0  = \omega ^2 f^{ - 1} \left( r \right) + \left( {2m^2 \beta  - 1} \right)m^2$. Solving above equation, one yields
\begin{equation}
\label{eq25}
{W_ \pm   =  \pm \int {\sqrt {f\left( r \right)m^2 \left( {1 - 2\beta m^2 } \right) + \omega ^2 } f^{ - 1} \left( r \right)\left[ {1 + \beta \left( {m^2  + f^{ - 1} \left( r \right)\omega ^2 } \right)} \right]dr,} }
\end{equation}
where   $P_4  = 2\beta f\left( r \right)^2$,   $P_2=f\left( r \right)\left( {4m^2 \beta  + \sqrt {8\beta \lambda }  - 1} \right)$ and   $P_0  = \omega ^2 f^{ - 1} \left( r \right) + m^2 \left[ {2\beta m^2  + \left( {\sqrt {8\beta \lambda }  - 1} \right)} \right] - \sqrt {{\lambda  \mathord{\left/ {\vphantom {\lambda  {2\beta }}} \right. \kern-\nulldelimiterspace} {2\beta }}}  + \lambda$. Neglecting the higher orders $\beta$ of and solving above equation, one yields
\begin{equation}
\label{eq25}
W_ \pm   = \pm \frac{1}{{f\left( r \right)}}\sqrt {f\left( r \right)\left( {m^2  - \lambda  + \sqrt {{\lambda  \mathord{\left/
 {\vphantom {\lambda  {2\beta }}} \right.
 \kern-\nulldelimiterspace} {2\beta }}} } \right) + \omega ^2 }   \left\{ {1 + \beta \left[ {m^2  + f^{ - 1} \left( r \right)\omega ^2 } \right] + \sqrt {{{\beta \lambda } \mathord{\left/
 {\vphantom {{\beta \lambda } 2}} \right.
 \kern-\nulldelimiterspace} 2}} } \right\}dr,
\end{equation}
where the ${ +  \mathord{\left/ {\vphantom { +   - }} \right. \kern-\nulldelimiterspace}  - }$ denote the outgoing/incoming solutions of emitted particles. In order to solve above equation, one needs to find the residue of Eq. (\ref{eq25}) on the event horizon. By expanding a Laurent series on the event horizon and keeping the first order term of  $\beta$, the result of Eq. (\ref{eq25}) takes on the form as
\begin{equation}
\label{eq26}
 W\left( {r_H } \right)_ \pm   =  \pm \frac{{i\pi r_H \omega }}{{D - 3}}\left\{ {1 + \sqrt {\frac{{\beta \lambda }}{8}}  + } \right.\beta  \left. {\left[ {\frac{{m^2  + \lambda }}{2} + \frac{{\left( {D - 2} \right)\omega ^2 }}{{\left( {D - 3} \right)}}} \right]} \right\} + \Delta \left( {realpart} \right).
\end{equation}
Because real part of Eq. (\ref{eq26}) is irrelevant to the tunneling rate, we only keep the imaginary part. For obtaining the tunneling rate from
Eq. (\ref{eq26}), one needs to solve the factor-two problem \cite{ch33,chvv1+}.
One of the best ways to solve this problem is to adopt the temporal contribution expression.
According to \cite{chv1+,chv2+,chv3+,chv4+,chv5+}, the spatial part to the tunneling rate of emitted particle is
\begin{equation}
\label{eq1v+}
\begin{array}{l}
 \Gamma  \propto \exp \left( { - {\mathop{\rm Im}\nolimits} \oint {p_r dr} } \right) = \exp \left[ {{\mathop{\rm Im}\nolimits} \left( {\int {p_r^{out} dr}  - \int {p_r^{in} dr} } \right)} \right] \\
 \\
 = \exp \left\{ { - \frac{{2\pi r_H \omega }}{{D - 3}}\left\{ {1 + \sqrt {\frac{{\beta \lambda }}{8}}  + \beta \left[ {\frac{{m^2  + \lambda }}{2} + \frac{{\left( {D - 2} \right)\omega ^2 }}{{\left( {D - 3} \right)}}} \right]} \right\}} \right\},
 \\
 \end{array}
\end{equation}
where $p_r  = \partial _r W$. However, as point in \cite{chv2+}, the authors showed that the temporal contribution to the tunneling amplitude was lost in the above discussion. For incorporating the temporal contribution into our calculation, we need use Kruskal coordinates $(T,R)$. The region exterior is given by
\begin{equation}
\label{eq2v+}
\begin{array}{*{20}c}
   {T = \exp \left( {\kappa r_* } \right)\sinh \left( {\kappa t} \right),} & {R = \exp \left( {\kappa r_* } \right)\cosh \left( {\kappa t} \right)}  \\
\end{array},
\end{equation}
where $r_*  = r + \frac{1}{{2\kappa }}\ln \frac{{r - r_H }}{{r_H }}$ is the tortoise coordinate and $\kappa$ is the surface gravity  of ST black hole. In order to connect the interior region and the exterior region across the horizon, one can rotate the time $t$ as $t\rightarrow t - {{i\pi } \mathord{\left/ {\vphantom {{i\pi } {2\kappa }}} \right. \kern-\nulldelimiterspace} {2\kappa }}$. By this operation, one obtains an additional imaginary contribution ${\mathop{\rm Im}\nolimits} \left( {\omega \Delta t^{out,in} } \right) = \omega {\pi  \mathord{\left/ {\vphantom {\pi  {2\kappa }}} \right.  \kern-\nulldelimiterspace} {2\kappa }}$. Therefore, the total temporal contribution becomes to ${\mathop{\rm Im}\nolimits} \omega \Delta t = \omega {\pi  \mathord{\left/ {\vphantom {\pi  \kappa }} \right. \kern-\nulldelimiterspace} \kappa }$. According to Eq. (\ref{eq1v+}), the GUP corrected tunneling rate of emitted particle across the horizon is derived as
\begin{equation}
\label{eq281}
 \Gamma  \propto \exp \left[ { - {\mathop{\rm Im}\nolimits} \left( {\omega t + {\mathop{\rm Im}\nolimits} \oint {p_r dr} } \right)} \right]
 = \exp \left\{ { - \frac{{4\pi r_H \omega }}{{D - 3}}\left\{ {1 + \sqrt {\frac{{\beta \lambda }}{8}}  + \beta \left[ {\frac{{m^2  + \lambda }}{2} + \frac{{\left( {D - 2} \right)\omega ^2 }}{{\left( {D - 3} \right)}}} \right]} \right\}} \right\},
\end{equation}
Employing the Boltzmann factor, the GUP corrected Hawking temperature is
\begin{equation}
\label{eq28}
T_H  = T_0 \left\{ {1 + \sqrt {\frac{{\beta \lambda }}{8}}  + \beta \left[ {\frac{1}{2}\left( {m^2  + \lambda } \right) + \frac{{\left( {D - 2} \right)\omega ^2 }}{{\left( {D - 3} \right)}}} \right]} \right\}^{ - 1} ,
\end{equation}
where  $T_0  = {{\left( {D - 3} \right)} \mathord{\left/ {\vphantom {{\left( {D - 3} \right)} {4\pi r_H }}} \right. \kern-\nulldelimiterspace} {4\pi r_H }}$  is the semi-classical Hawking temperature of the ST black hole. Now, turn to calculate the entropy of ST black hole. Base on the first law of black hole thermodynamics, the entropy can be expressed as
\begin{equation}
\label{eq29}
 S = \int {T_H^{ - 1} dM}  = \int {\frac{{4\pi r_H }}{{D - 3}}\left[ {\frac{{\left( {D - 2} \right)\varpi }}{{16\pi GM}}} \right]^{\frac{1}{{D - 3}}} }  \left\{ {1 + \sqrt {\frac{{\beta \lambda }}{8}}  + \beta \left[ {\frac{1}{2}\left( {m^2  + \lambda } \right) + \frac{{\left( {D - 2} \right)\omega ^2 }}{{\left( {D - 3} \right)}}} \right]} \right\}dM.
\end{equation}
The above equation cannot be evaluated exactly for general  $D$. According to the standard Hawing radiation theory, all particles near the event horizon are seem effectively massless. Therefore, we do not consider the mass of emitted particles in the following discussion.

\section{Remnants of ST black hole}
\begin{figure}
\centering
\subfigure{
\begin{minipage}[b]{0.3\textwidth}
\includegraphics[width=1\textwidth]{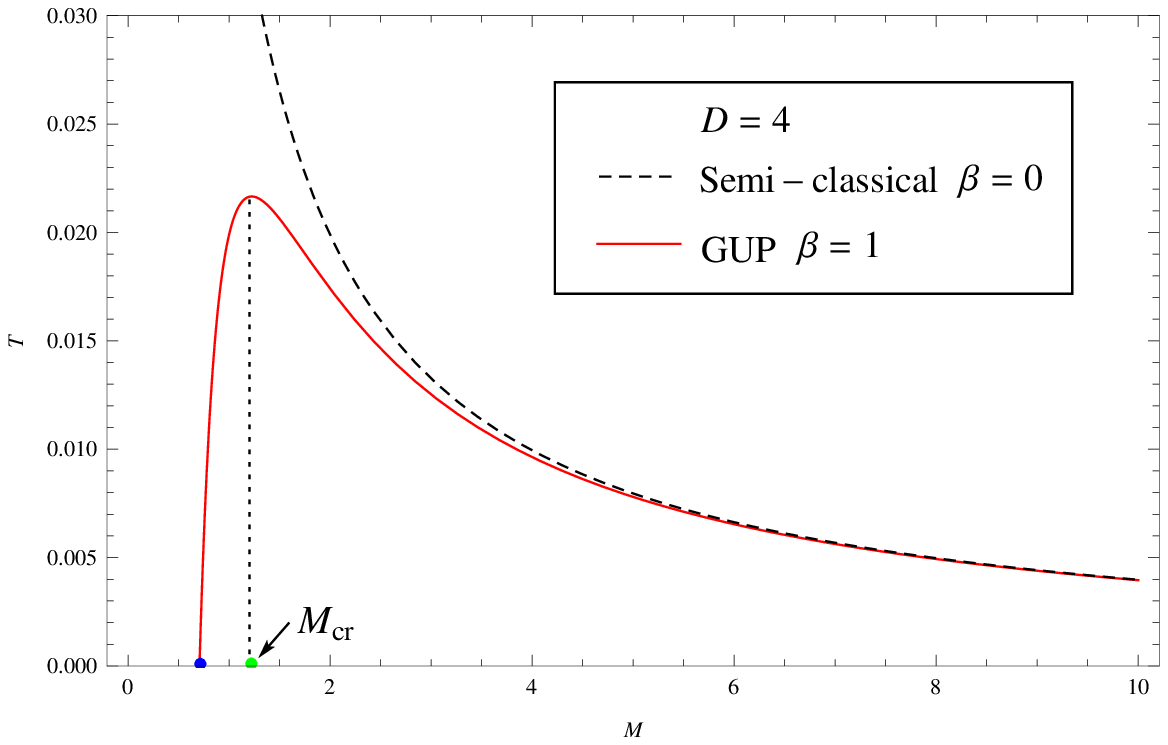}
\end{minipage}
}
\subfigure{
\begin{minipage}[b]{0.3\textwidth}
\includegraphics[width=1\textwidth]{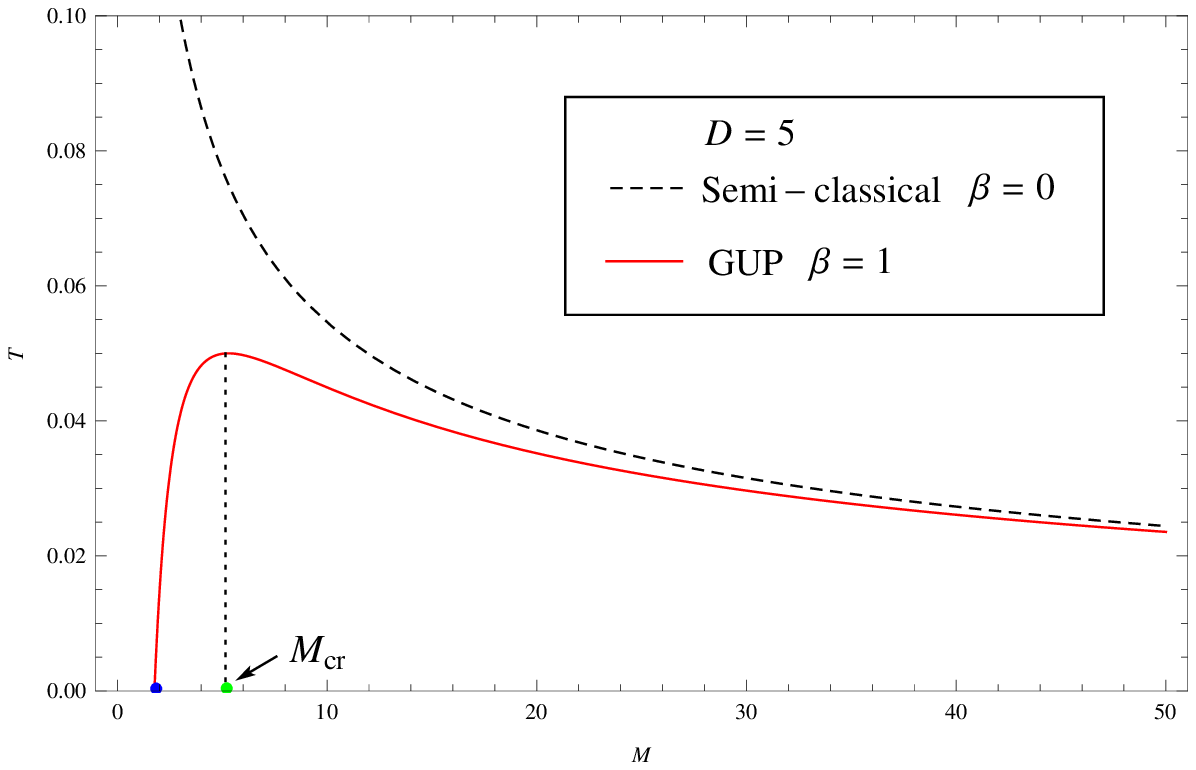}
\end{minipage}
}
\subfigure{
\begin{minipage}[b]{0.3\textwidth}
\includegraphics[width=1\textwidth]{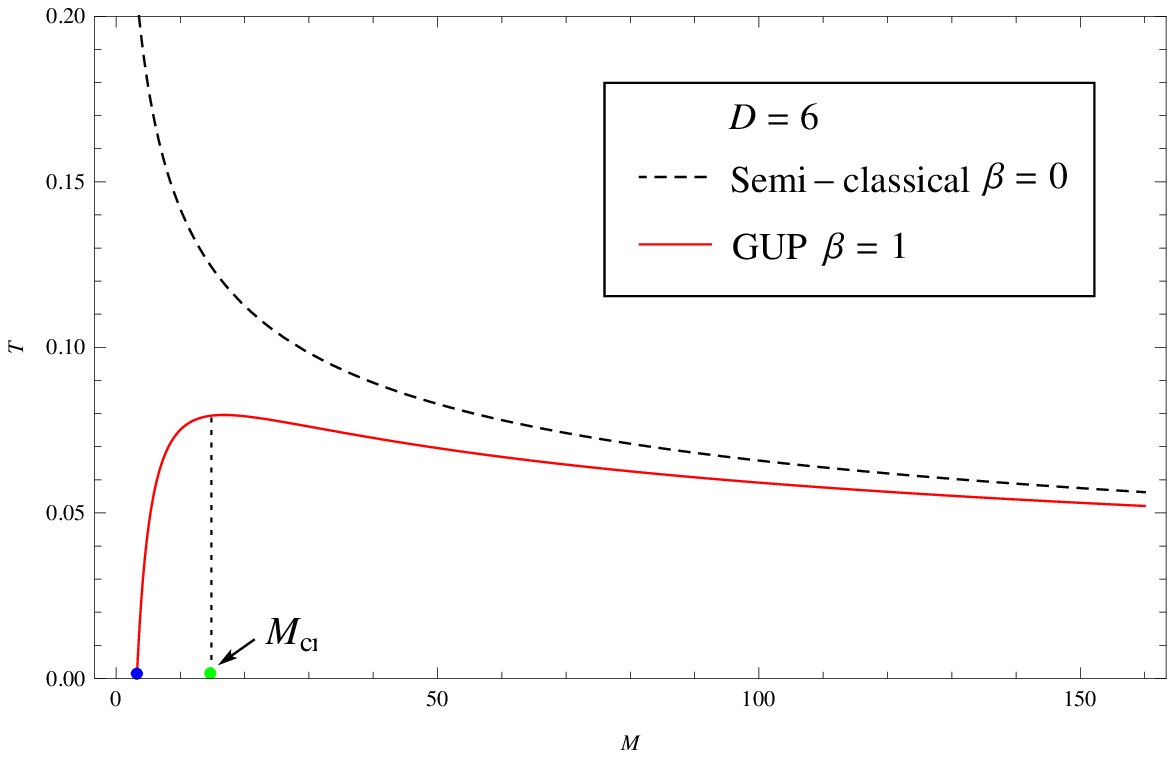}
\end{minipage}
}
\caption{Semi-classical and GUP corrected Hawking temperature of ST black hole for different values of mass. We set $M_P  = c = \hbar = 1$, $D = 4,5$ and $6$  for simplicity.}
\label{fig1}       
\end{figure}

A lot of work showed that the GUP can leads to the black hole remnant \cite{ch17+,ch19,ch32b+,ch19+,ch21,ch1c,ch18+,ch27,ch28,ch28+,ch29+,ch20+,ch20b+}.
Therefore, it is interesting to investigate the remnant of ST black hole. According to the saturated form of the uncertainty principle,
one gets a lower bound on the energy of the emitted particle in Hawking radiation, which can be expressed as \cite{ch18,ch34}
\begin{equation}
\label{eq30}
\omega  \ge {\hbar  \mathord{\left/
 {\vphantom {\hbar  {\Delta x}}} \right.
 \kern-\nulldelimiterspace} {\Delta x}}.
\end{equation}
 Near the event horizon of ST black hole, it is possible to take the value of the uncertainty in position as the radius of the black hole, that is \cite{ch24,ch25}
\begin{equation}
\label{eq31}
\Delta x \approx r_{BH} =  r_H.
\end{equation}
Putting Eq. (\ref{eq30}) and Eq. (\ref{eq31}) into Eq. (\ref{eq28}), and expanding, one has
\begin{equation}
\label{eq32}
 T_H  =T_0 \left\{ {1 + \frac{3}{2}\sqrt {\frac{{\beta \lambda }}{2}}  + \beta \left[ {\frac{{\left( {D - 2} \right)\omega ^2 }}{{\left( {D - 3} \right)}} - \frac{\lambda }{2}} \right]} \right\}^{ - 1}  \simeq T_0 \left\{ {\frac{{2\left[ {4\left( {D - 2} \right)\hbar ^2 \beta  + \left( {D - 3} \right)r_H^2 \left( {\sqrt {2\beta \lambda }  + 2\beta \lambda  - 4} \right)} \right]}}{{r_H^2 \left( {D - 3} \right)\left( {\beta \lambda  - 8} \right)}}} \right\}.
\end{equation}
It is clear that  $T_{H}$ sensitively depends  on the event horizon of ST black hole, the spacetime dimension $D$, the angular momentum of emitted particles
and the quantum gravity effect  $\beta$. An important relation should be mentioned,
 when $r_H  < \sqrt {\frac{{4\left( {D - 2} \right)\beta \hbar ^2 }}{{\left( {D - 3} \right)\left( {4 - 2\beta \lambda  - \sqrt {2\beta \lambda } } \right)}}}$,
 the Hawking temperature goes to negative, it violates the laws of black hole thermodynamics and has no physical meaning.
  Therefore, this relation indicates the existence of a minimum radius, where the Hawking temperature equals to zero,
   that is,
\begin{equation}
\label{eq33xx}
r_{\min }  = \sqrt {\frac{{4\left( {D - 2} \right)\beta \hbar ^2 }}{{\left( {D - 3} \right)\left( {4 - 2\beta \lambda  - \sqrt {2\beta \lambda } } \right)}}} = \ell _p \sqrt {\frac{{4\hbar ^2 \left( {D - 2} \right)\beta _0 }}{{\left( {D - 3} \right)\left( {4\hbar ^2  - 2\lambda \beta _0 \ell _p^2  - \ell _p \hbar \sqrt {2\lambda \beta _0 } } \right)}}}.
\end{equation}
In addition, we can also express  Eq. (\ref{eq32}) in terms of the mass of the ST black hole to obtain the temperature-mass relation
\begin{equation}
\label{eq33}
T_H  \simeq \frac{{D - 3}}{{4\pi }}\left[ {\frac{{\left( {D - 2} \right)\varpi _{D - 2} }}{{16\pi GM}}} \right]^{\frac{1}{{D - 3}}} {\frac{{2\left\{ {4\hbar ^2 \beta \left( {D - 2} \right) - \left( {D - 3} \right)\left[ {\frac{{\left( {D - 2} \right)\varpi _{D - 2} }}{{16\pi GM}}} \right]^{\frac{2}{{D - 3}}} \left( {4 - \sqrt {2\beta \lambda }  - 2\beta \lambda } \right)} \right\}}}{{\left( {D - 3} \right)\left[ {\frac{{\left( {D - 2} \right)\varpi _{D - 2} }}{{16\pi GM}}} \right]^{\frac{2}{{D - 3}}} \left( {\beta \lambda  - 8} \right)}}} .
\end{equation}
From Eq. (\ref{eq33}), we find  that the GUP corrected temperature has physical meaning as far as the mass of ST black hole satisfies inequality $M \ge \frac{{\left( {D - 2} \right)\varpi _{D - 2} }}{{16\pi G}}[\frac{{4\left( {D - 2} \right)\hbar ^2 \beta }}{{\left( {D - 3} \right)\left( {4 - 2\beta \lambda  - \sqrt {2\beta \lambda } } \right)}}]^{\frac{{D - 3}}{2}}$, which implies that the mass of ST black hole has a minimum value
\begin{equation}
\label{eq34}
M_{\min }  = \frac{{\left( {D - 2} \right)\varpi _{D - 2} }}{{16\pi G}}\left[ {\frac{{4\left( {D - 2} \right)\hbar ^2 \beta }}{{\left( {D - 3} \right)\left( {4 - 2\beta \lambda  - \sqrt {2\beta \lambda } } \right)}}} \right]^{\frac{{D - 3}}{2}} = \frac{{\left( {D - 2} \right)M_p }}{{8\Gamma \left( {\frac{{D - 1}}{2}} \right)}}\left[ {\frac{{4\pi \beta _0 \hbar ^2 \left( {D - 2} \right)}}{{c^2 \left( {D - 3} \right)\left( {4 - \frac{{2\lambda \beta _0 }}{{M_p^2 c^2 }} - \sqrt {\frac{{2\lambda \beta _0 }}{{M_p^2 c^2 }}} } \right)}}} \right]^{\frac{{D - 3}}{2}}
\end{equation}
Obviously, the minimum mass is related to the Plank mass. According to Eq. (\ref{eq33}) and Eq. (\ref{eq34}), the behaviors of GUP corrected Hawking temperature and original Hawking temperature of ST black hole are plotted in Fig. \ref{fig1}.

In Fig. \ref{fig1}, the dash  black lines and solid red lines in the diagrams illustrate the original Hawking temperature and GUP corrected temperature of
the ST black hole. It is easy to find that the GUP corrected temperature is lower than  original Hawking temperature. Besides, different values of  $D$ give similar behavior of Hawking temperature. For large mass of black hole, the GUP corrected temperature tends to the
original Hawking temperature value because the effect of quantum gravity is negligible at that scale. However, as mass of black hole decreases,
the GUP corrected temperature reaches the maximum value (at the critical mass  $M_{cr}$, which is marked by green dot), and then decreases to zero when mass
approaches to minimum value of mass ($M_{min}\sim M _p$ , which is marked by blue dot). The GUP corrected temperature is unphysical below the $M_{min}$,
it signals the existence of a black hole remnant $M_{res}=M_{min}$. The black hole remnant can be further confirmed from the heat capacity.

Since the thermodynamic stability of black hole is determined by the heat capacity $\mathcal{C}$,
a further inspection in the existence of the black hole remnant can be made by investigating the heat capacity of the ST black hole. The GUP corrected heat capacity is given by
\begin{equation}
\label{eq35}
\mathcal{C }= T_H \left( {\frac{{\partial S}}{{\partial T_H }}} \right) = T_H \left( {\frac{{\partial S}}{{\partial M}}} \right)\left( {\frac{{\partial T_H }}{{\partial M}}} \right)^{ - 1}  = \frac{{\cal A}}{{\cal B}}.
\end{equation}
According to Eq. (\ref{eq30}) and Eq. (\ref{eq31}), the entropy can rewritten as
\begin{equation}
\label{eq36}
 S = \int {\frac{{4\pi r_H }}{{D - 3}}\left[ {\frac{{\left( {D - 2} \right)\varpi }}{{16\pi GM}}} \right]^{\frac{1}{{D - 3}}} \left\{ {1 + \sqrt {\frac{{\beta \lambda }}{8}}  + } \right.} \left.\beta {\left\{ {\frac{\lambda }{2} + \frac{{\left( {D - 2} \right)\omega ^2 \hbar ^2 }}{{\left( {D - 3} \right)}}\left[ {\frac{{\left( {D - 2} \right)\varpi }}{{16\pi GM}}} \right]^{\frac{2}{{D - 3}}} } \right\}} \right\}dM. \\
\end{equation}
and the $\mathcal{A}$  and  $\mathcal{B}$ in Eq. (\ref{eq35}) are defined by
\begin{equation}
\label{eq37}
\begin{array}{l}
 {\cal A} = 2^{2 + \frac{4}{{D - 3}}} \left[ {\frac{G}{{\left( {D - 2} \right)\varpi }}} \right]^{\frac{1}{{D - 3}}} \left( {M\pi } \right)^{1 + \frac{1}{{D - 3}}} \left( {4\hbar ^2 \beta \frac{{\left( {D - 2} \right)}}{{\left( {D - 3} \right)}} - \left[ {\frac{{16\pi GM}}{{\left( {D - 2} \right)\varpi }}} \right]^{\frac{2}{{D - 3}}} \left( {4 - \sqrt {2\beta \lambda }  - 2\beta \lambda } \right)} \right) \times  \\
 \\
 \left[ {\left( {1 + \frac{{\sqrt {2\beta \lambda } }}{4} + \frac{{\beta \lambda }}{2}} \right) + \frac{{\hbar ^2 \beta \left( {D - 2} \right)^{1 + \frac{2}{{D - 3}}} }}{{D - 3}}\left( {\frac{{16\pi GM}}{\varpi }} \right)^{\frac{2}{{D - 3}}} } \right] \\
 \end{array}
\end{equation}
\begin{equation}
\label{eq38}
 {\cal B} =  -\left\{ {\frac{{12\hbar ^2 \beta \left( {D - 2} \right)}}{{D - 3}} + \left( {2^{\frac{{D + 3}}{{D - 3}}}  - 3 \times 2^{\frac{6}{{D - 3}}} } \right)} \right. \left. { \left[ {\frac{{2GM\pi }}{{\left( {D - 2} \right)\varpi }}} \right]^{\frac{2}{{D - 3}}} \left( {4 - \sqrt {2\beta \lambda }  - 2\beta \lambda } \right)} \right\} \\
\end{equation}
Assuming $\beta=0$, one obtains the original specific heat of ST black hole from Eq. (\ref{eq35}). We find that specific heat  goes to zero at $
M  = \frac{{\left( {D - 2} \right)\varpi _{D - 2} }}{{16\pi G}}\left[ {\frac{{4\left( {D - 2} \right)\hbar ^2 \beta }}{{\left( {D - 3} \right)\left( {4 - 2\beta \lambda  - \sqrt {2\beta \lambda } } \right)}}} \right]^{\frac{{D - 3}}{2}}
$, which is equal to  $M_{min}$ from Eq. (\ref{eq34}). The behaviors of the heat capacity of ST black hole for  $D=4,5$ and $6$  are shown in Fig. 2.
\begin{figure}
\centering
\subfigure{
\begin{minipage}[b]{0.3\textwidth}
\includegraphics[width=1\textwidth]{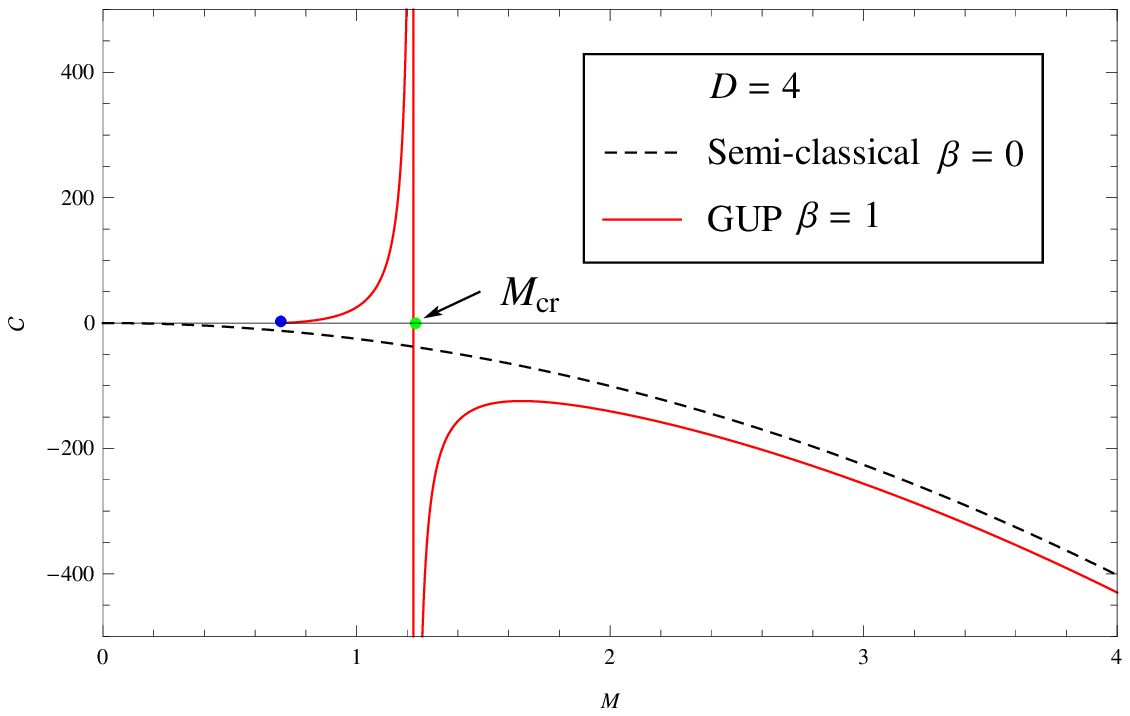}
\end{minipage}
}
\subfigure{
\begin{minipage}[b]{0.3\textwidth}
\includegraphics[width=1\textwidth]{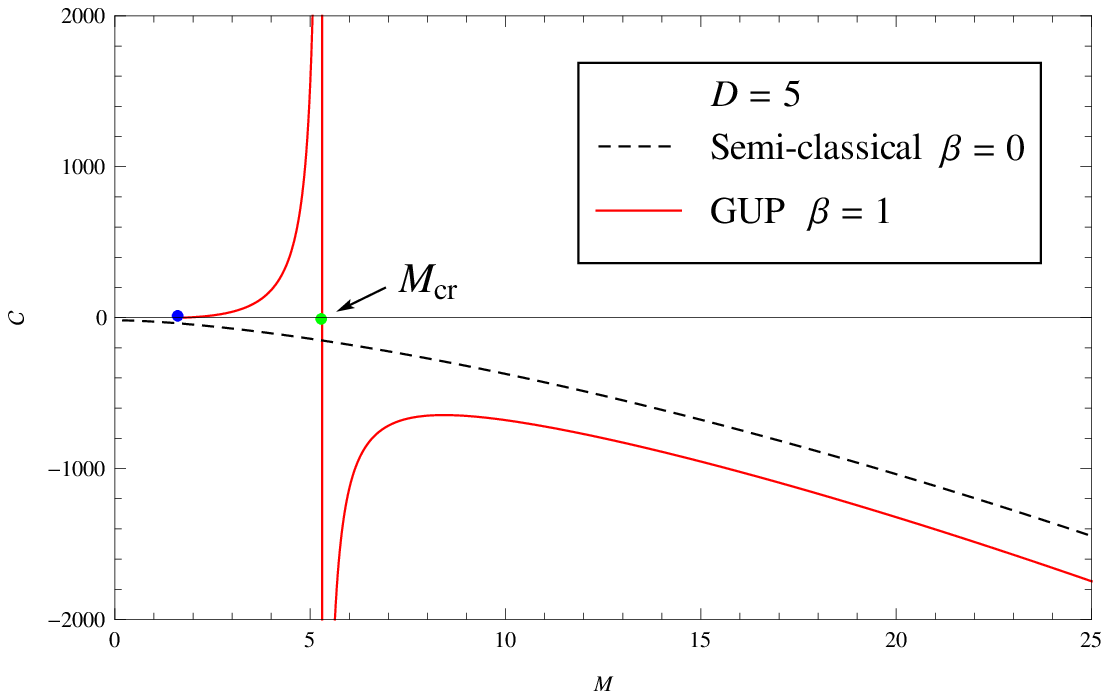}
\end{minipage}
}
\subfigure{
\begin{minipage}[b]{0.3\textwidth}
\includegraphics[width=1\textwidth]{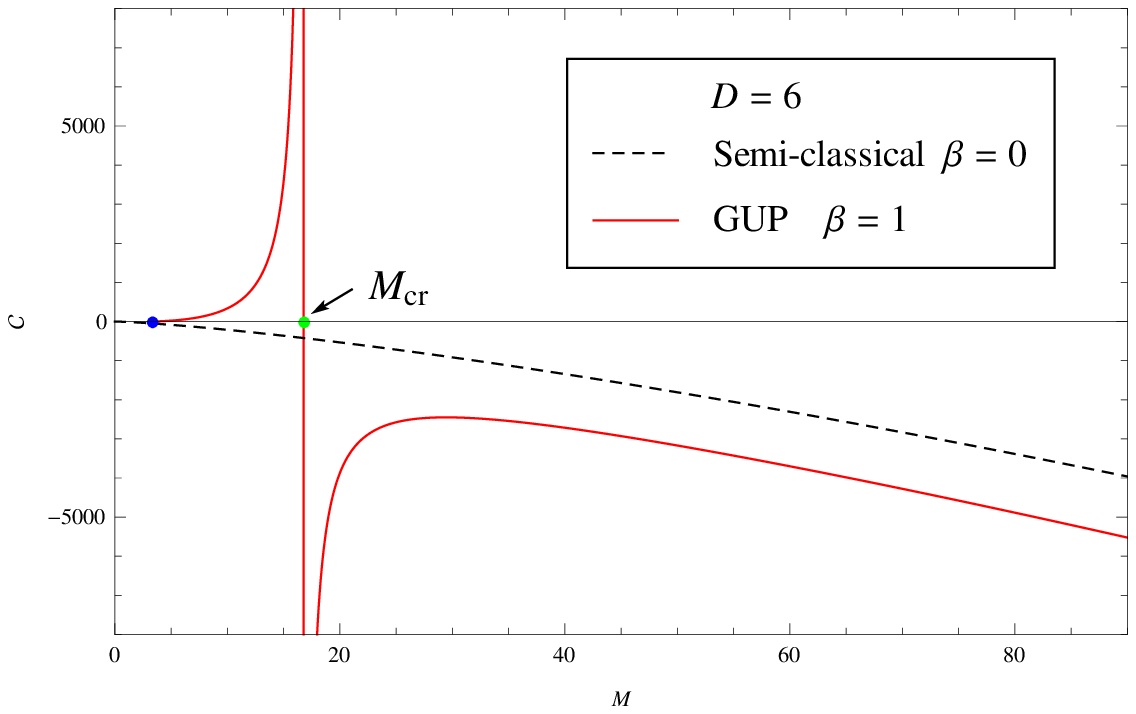}
\end{minipage}
}
\caption{Semi-classical and GUP corrected specific heat of ST black hole for different values of mass. We assumed $M_P  = c = \hbar = 1$,  $D=4, 5$ and  $6$.}
\label{fig2}       
\end{figure}

In Fig. \ref{fig2}, one can see the specific heat versus the mass of ST black hole.
Notably, the different values of  $D$ give similar behavior of specific heat.
The black dash lines correspond the original specific heat, they are negative values and go to zero when  $M\rightarrow0$.
The GUP corrected specific heat is represented by red solid lines. It is clear that the GUP corrected specific heat diverges at green dot,
where the GUP corrected temperature reaches its maximum value $M_{cr}$. When the mass of black hole is enough large, the behavior of GUP corrected
specific heat is similar as the original case. By decreasing the mass of ST black hole, the GUP corrected specific heat becomes smaller and
departs from the original ST black hole behaviour. However, at $M = M_{cr}$, the GUP corrected specific heat has a vertical asymptote and becomes
 to position value, it implies a thermodynamic phase transition happened from  $\mathcal{C}<0$ (unstable phase) to  $\mathcal{C}>0$ (stable phase),
 this phase transition is also found in the GUP black holes \cite{ch1n+} and the framework of gravity's rainbow \cite{ch8,ch9,ch10}.
 Finally, the GUP corrected specific heat decreases to the zero as mass decreases to  $M_{min}$ (blue dot).
 The $\mathcal{C}=0$  means that the black hole cannot exchange its energy with environment,
 hence the GUP stops the evolution of black holes at this point and leads to the black hole remnant, that is, $M_{\min }  = M_{res}$.

\section{Black hole remnants in the colliders}
The production of black holes at the colliders such as LHC is one of the most exciting predictions of physics.
Due to the Eq. (\ref{eq34}), one can calculate whether the black holes could be formed at the LHC.
the minimum energy needed to form black hole in a collider is given by
\begin{equation}
\label{eq377}
E_{\min }^{GUP}  = \frac{{\left( {D - 2} \right)M_p }}{{8\Gamma \left( {\frac{{D - 1}}{2}} \right)}}\left[ {\frac{{4\pi \beta _0 \hbar ^2 \left( {D - 2} \right)}}{{c^2 \left( {D - 3} \right)\left( {4 - \frac{{2\lambda \beta _0 }}{{M_p^2 c^2 }} - \sqrt {\frac{{2\lambda \beta _0 }}{{M_p^2 c^2 }}} } \right)}}} \right]^{\frac{{D - 3}}{2}}.
 \end{equation}
In order to investigate the minimal energy for black hole formation,
we use the latest observed limits on the ADD model \cite{ch2c} parameter $M_p$ with next-to-leading-order (NLO) $K$-factor \cite{ch34d+,ch345d+}.
When setting $\beta_0=c=\hbar=1$ and $\lambda=0.001$, the minimum energy to form black hole $E_{\min }^{{{\rm{GUP}}}}$ is shown in Tab. \ref{tab1}.
\begin{table}[htbp]
\caption {\label{tab1} The latest experimental limits on $M_p$, the minimal energy for black hole formation $E^{\rm GUP}_{\min }$ and $E_{\min }^{{\rm{GR}}}$
in different dimensions $D$.}
\centering
\begin{tabular}{c c c c c}
\toprule
$D$ &  $M_p$ &   $E^{\rm GUP}_{\min }$ &  $E_{\min }^{{\rm{GR}}}$ \cite{ch345d+}\\
\midrule
6&  4.54 TeV&  14.6 TeV&  9.5 TeV\\
7&  3.51 TeV&  17.0 TeV&  10.8 TeV\\
8&  2.98 TeV&  18.7 TeV&  11.8 TeV\\
9&  2.71 TeV&  19.7 TeV&  12.3 TeV\\
10& 2.51 TeV&  19.2 TeV&  11.9 TeV\\
\bottomrule
\end{tabular}
\end{table}

We also compare our results with the results obtained in Gravity's Rainbow (GR)
 $E_{\min }^{{\rm{GR}}}  = \frac{{\left( {D - 2} \right)}}{{8\Gamma \left( {\frac{{D - 1}}{2}} \right)}}\pi ^{\frac{{D - 3}}{2}} \eta ^{\frac{{D - 3}}{n}} M_p$,
 where $\eta (=1)$ and $n (=2)$ represent rainbow parameter and an integer, $M_p$ is the Planck mass \cite{ch345d+}.
 It is shown that our results are higher than $E_{\min }^{{\rm{GR}}}$. This difference is caused by different modified gravity theories. Quite recently, the protons collided in the LHC has reached a new energy  regime at 13 TeV \cite{ch345f+}, but it is still smaller than the $E^{\rm GUP}_{\min }$ in $D=6$, which implies the black hole can not be produced in the LHC. This may explant the absence of black holes in current LHC.

 Moreover, we only fix $\beta_0=1$ in Tab. \ref{tab1}. However, from the expression of $E_{\min }$, we find it is closely related to the dimensionless constant $\beta_0$, which indicates that the different value of $\beta_0$ may leads different value of minimum energy for black hole formation. The low bound of $\beta_0$ can be studied by the following formal
\begin{equation}
\label{eq39}
 \beta _0  > 4\chi  - \frac{{2\chi ^3 \lambda ^2 }}{{\left( {c^2 M_p^2  + 2\lambda \chi } \right)^2 }} - \frac{{7\lambda \chi ^2 }}{{c^2 M_p^2  + 2\lambda \chi }} - \frac{{c^2 M_p^2 \chi \sqrt {\chi \lambda \left( {8c^2 M_p^2  + 17\chi \lambda } \right)} }}{{\left( {c^2 M_p^2  + 2\chi \lambda } \right)^2 }}, \\
\end{equation}
where $\chi  = \frac{{c^2 \left( {D - 3} \right)}}{{4\pi \hbar ^2 \left( {D - 2} \right)}}\left[ {\frac{{13{\rm{TeV}} \times 8\Gamma \left( {\frac{{D - 1}}{2}} \right)}}{{\left( {D - 2} \right)M_p }}} \right]^{\frac{2}{{D - 3}}}$. The bounds on $\beta_0$ for $D=6,7,8,9,10$ are given in Tab. \ref{tab2}. Combining our results with earlier versions of GUP and some phenomenological implications in \cite{ch14,ch5c,ch7c,ch8c}, it indicates that  $\beta_0\sim1$.
\begin{table}[htbp]
\caption {\label{tab2}The lower bonds on $\beta_0$ for different $D$, we set $c=\hbar=1$ and $\lambda=0.001$.}
\centering
\begin{tabular}{c c c c c c}
\toprule
$D$ &  $6$  &   $7$ &   $8$&   $9$ &   $10$\\
\midrule
$\beta_0$&  0.9216&  0.8740&  0.8642&   0.8707&   0.8944\\
\bottomrule
\end{tabular}
\end{table}

\section{Conclusions}
In this work, we have investigated the GUP effect on the thermodynamics of ST black hole. First of all, we derived the modified Hamilton-Jacobi equation
 by employing the GUP with a quadratic term in momentum. With the help of modified Hamilton-Jacobi equation,
 the quantum tunneling from ST black hole has been studied. Finally, we obtained the GUP corrected Hawking temperature,
 entropy and heat capacity. For the original Hawing radiation, the Hawking temperature of ST black hole is related to its mass. However, our results showed that
if the effect quantum gravity is considered, the behavior of tunneling particle on the event is different from the original case, and the GUP corrected thermodynamic quantities are not only sensitively depended on the mass $M$ and the spacetime dimension $D$ of ST black hole, but also on the angular momentum parameter $\lambda$ and the quantum gravity term  $\beta$. Besides, we found that the GUP corrected Hawking temperature is smaller than the original case, it goes to zero when the mass of ST black hole reaches the minimal value $M_{min}$, which is the order of Planck scale,
it predicts the existence of a black hole remnant. For confirming the black hole remnant, the GUP corrected heat capacity has also been analyzed.
   It was shown that the GUP corrected heat capacity has a phase transition at  $M_{cr}$, where the GUP corrected temperature reaches its maximum value,
   then the GUP corrected heat vanishes when the mass approaches to  $M_{min}$ in the final stages of black hole evaporation.
   At this point, the ST black hole does not exchange the energy with the environment, hence the remnant of ST black hole is produced.
   The reason for this remnant is related to the fact that the quantum gravity effect is running as the size of black hole approaches to the Planck scale.
   The existence of black hole remnant implies that black holes would not evaporate, its information and singularity are enclosed in the event horizon.
   At last, we discussed the minimum energy to form black hole in the LHC.
   The results showed that the minimum energy to form black hole in our work is larger than the current energy scales of LHC,
   this may explains why people can not observe the black hole in the LHC.
   Our results are support by the results obtained in the framework of gravity's rainbow \cite{ch8,ch9,ch10}.
   Therefore, we think that the GUP effect can effectively prevents black hole from evaporating completely, and may solve the information loss
   and naked singularities problems of black hole \cite{ch35,ch36}.

\vspace*{3.0ex}
{\bf Acknowledgements}
\vspace*{1.0ex}

This work is supported by the Natural Science Foundation of China (Grant No. 11573022).

\end{document}